\documentclass{UApreprint}
\usepackage{epsfig}
\usepackage{graphics}
\usepackage{fullpage}
\usepackage{verbatim}
\usepackage[tight]{subfigure}
\usepackage{amssymb}
\newcommand{\be}{\begin{equation}}
\newcommand{\ee}{\end{equation}}
\newcommand{\bea}{\begin{eqnarray}}
\newcommand{\eea}{\end{eqnarray}}
\newcommand{\bean}{\begin{eqnarray*}}
\newcommand{\eean}{\end{eqnarray*}}
\newcommand{\bi}{\begin{itemize}}
\newcommand{\ei}{\end{itemize}}
\newcommand{\bdm}{\begin{displaymath}}
\newcommand{\edm}{\end{displaymath}}

\newcommand{\meg}{{\mu\to e\gamma}}

\newcommand{\la}{\mathcal{L}}

\newcommand{\Ye}{\mathbf{Y}_e}
\newcommand{\Yn}{\mathbf{Y}_\nu}

\newcommand{\Ae}{\mathbf{A}_e}
\newcommand{\An}{\mathbf{A}_\nu}

\newcommand{\OL}{\mathbf{O}_\mathrm{L}}
\newcommand{\OR}{\mathbf{O}_\mathrm{R}}

\newcommand{\ON}{\mathbf{O}_\mathrm{ne}}
\newcommand{\cc}{\;+\;\mathrm{c.\;c.}}

\newcommand{\Uf}{\mathbf{U}_\mathrm{\tilde{f}}}
\newcommand{\Un}{\mathbf{U}_\mathrm{\tilde{n}}}

\begin{document}

\title{Hidden-Sector Dynamics and the Supersymmetric Seesaw}
\author{Bruce A. Campbell~$^{1,2}$, John Ellis~$^2$, and David W. Maybury~$^1$}
\address{$^1$Department of Physics, Carleton University,
1125 Colonel By Drive, \\ Ottawa ON K1S 5B6, Canada \\
$^2$ Theory Division, PH Department,
CERN, CH-1211 Geneva 23, Switzerland}

\date{October 27th, 2008}

\abstract{In light of recent analyses that have shown that nontrivial hidden-sector dynamics in 
models of supersymmetry breaking can lead to a significant impact on the predicted 
low-energy supersymmetric spectrum, we extend these studies to consider hidden-sector effects 
in extensions of the MSSM to include a seesaw model for neutrino masses. A dynamical hidden 
sector in an interval of mass scales below the seesaw scale would yield renormalization-group 
running involving both the anomalous dimension from the hidden sector and the seesaw-extended MSSM renormalization group equations (RGEs). These effects interfere in general, 
altering the generational mixing of the sleptons, and allowing for a substantial change to the 
expected level of charged-lepton flavour violation in seesaw-extended MSSM models. These results 
provide further support for recent theoretical observations that  knowledge of the hidden sector is required in order to make concrete low-energy predictions, if the hidden sector is strongly coupled. 
In particular, hidden-sector dynamics may impact our ability to reconstruct 
the supersymmetric seesaw parameters from low-energy observations.}

\archive{}
\preprintone{CERN-PH-TH/2008-210}
\preprinttwo{}

\submit{}

\maketitle

\section{Introduction}

Weak-scale supersymmetry in the guise of the minimal supersymmetric extension of the
Standard Model 
(MSSM)~\cite{Nilles:1983ge,Haber:1984rc} provides an elegant solution to the gauge hierarchy 
problem, satisfies electroweak precision constraints, and predicts weakly-interacting dark matter 
(see, for example, \cite{Ellis:2007fu}). However, the MSSM requires additional physics responsible 
for breaking supersymmetry itself (for reviews see~\cite{Drees:2004jm,Ramond:1999vh,Terning:2006bq,Weinberg:2000cr}). This additional physics is expected to include a hidden sector that 
breaks supersymmetry spontaneously, and a messenger sector, which communicates the symmetry 
breakdown to the visible sector fields of the MSSM. The MSSM predicts 
gauge-coupling-constant unification~\cite{Dimopoulos:1981zb,Georgi:1974yf,Dimopoulos:1981yj} 
at $M_{\mathrm{U}}\sim 2\times 10^{16}$~GeV 
if there is a desert between the weak scale and the unification scale. The success of this
prediction encourages the hope that one may gain insights into 
high-scale physics through the renormalization group equations (RGEs) of the MSSM. 

The observation of neutrino oscillations~\cite{Cleveland:1998nv,Ahmad:2001an,Ahmad:2002jz,Ahmad:2002ka,Fukuda:1998mi,Fukuda:1998ah,Fukuda:2000np,Fukuda:2001nk,Eguchi:2002dm,Araki:2004mb,Ahn:2002up,Aliu:2004sq,Hampel:1998xg,Anselmann:1995ag,Abdurashitov:2002nt,Abdurashitov:1999bv,Abdurashitov:1994bc} suggestions the existence of new high-scale physics responsible for 
the small neutrino masses inferred from experiment. A popular framework for generating small
neutrino masses invokes heavy Majorana gauge-singlet fermions that, once integrated out of the 
theory, yield the dimension-five operator $LLHH$ with a coefficient suppressed by a factor of
the heavy gauge-singlet mass scale. This framework -- called the seesaw  mechanism (for a
review, see~\cite{Yan}) --  in its simplest incarnation (the Type-I seesaw), demands that the Majorana 
mass scale appears below the unification scale, in order to satisfy the constraints provided by 
neutrino oscillations, unitary, and 
triviality~\cite{Campbell:2006qk,Casas:1999ac,Antusch:2001ck,Antusch:2003kp}, and typical
values of the Majorana scale are near $M_R \sim 10^{14}$ GeV. 

Since the seesaw mechanism 
violates lepton number by two units, its natural supersymmetric extension is consistent with 
R-parity conservation, preserving the stability of the lightest supersymmetric particle and 
hence its candidacy for cold dark matter. The supersymmetric seesaw can be
therefore be incorporated as a simple extension of the MSSM. The appearance of the 
Majorana scale in the desert below 
the unification scale implies an interval in which the RGEs of the MSSM are augmented by the 
interactions contributing to the seesaw mechanism. The change in the renormalization-group 
flow of the MSSM induced by the heavy gauge-singlet Majorana fermions does not affect the 
RGE evolution of the gauge couplings at leading order, but does, in general, generate
off-diagonal mixing in the slepton sector that can lead to a observable predictions for 
charged-lepton flavour-violating decays. This prospect offers further insight into high-scale physics 
from low-energy observations.

While it has been long understood that different means of supersymmetry breaking lead to 
different sparticle spectra and different low-energy phenomenologies~\cite{Nilles:1983ge,VanNieuwenhuizen:1981ae,Dine:1981za,Dimopoulos:1981au,Nappi:1982hm,Barbieri:1982eh,AlvarezGaume:1981wy,Dimopoulos:1982gm,Dine:1993yw,Dine:1994vc,Dine:1995ag,Giudice:1998bp,Choi:2007ka}, only recently has it been realized that hidden-sector dynamics may play an 
important role in low-energy predictions~\cite{Dine:2004dv,Cohen:2006qc,Murayama:2007ge,Schmaltz:2006qs}. Specifically, it has been shown that if the hidden sector contains strong self-interactions, 
hidden-sector renormalization effects can influence significantly the renormalization-group 
running of the MSSM scalar sector. These hidden-sector renormalization effects may modify
in an observable way the simple mass relations expected naively from scalar-mass unification at the 
unification scale.

Observable-sector effects from the hidden sector result from quantum corrections
that correct the non-renormalizable operators introduced at the messenger scale required 
to communicate supersymmetry breaking to the observable sector. Since the hidden-sector scale 
sits in the desert below the unification scale (e.g., $M_{\mathrm{hidden}}\sim 10^{12}$ GeV in 
typical gravity-mediated models), hidden-sector renormalization effects will be present alongside the 
seesaw mechanism. Furthermore, since the hidden-sector renormalization affects directly
the diagonal scalar-mass terms, the effect will alter the ratio between the radiatively-generated 
seesaw off-diagonal slepton masses and the diagonal terms. In general, models with moderately 
strongly-coupled hidden sectors will therefore have an impact on the expected amount of 
charged-lepton flavour violation.

This paper is a continuation of our previous work~\cite{CEM1} in which we explored
the possible measurability of the diagonal scalar-mass effects. Here we explore the effects of 
hidden-sector renormalization on the seesaw extension of the MSSM with 
gravity-mediated supersymmetry breaking. We examine the consequent impact of hidden-sector 
effects on charged-lepton flavour violation, in particular on the induced rate for $\meg$ in the seesaw extension of the MSSM, using model-independent anomalous-dimension parametrizations of the 
hidden sector's renormalization group evolution. In this fashion, we examine a wide range of 
model classes for the hidden sector and their effect on the predictions for charged-lepton flavour 
violation in models with a supersymmetric seesaw. We also discuss the impact of these effects on 
the possibility of reconstructing the supersymmetric seesaw parameters from weak-scale observations.

\section{The MSSM Seesaw}
\label{LFV_MSSM}

We consider the class of gravity-mediated supersymmetry-breaking models that lead to the 
constrained MSSM (CMSSM) 
at the unification scale, with universal flavour-diagonal soft masses, universal gaugino masses, and 
tri-linear A-terms proportional to the superpotential Yukawa couplings and the universal scalar mass. Charged-lepton flavour violation in supersymmetric seesaw models arises from 
renormalization-group running of the seesaw sector between the unification scale and the Majorana 
mass scale~\cite{Borzumati:1986qx,Hisano:1995cp}. The seesaw sector thereby induces
radiatively off-diagonal slepton mass terms that contribute to flavour-violating decays. 

There is an elegant parametrization~\cite{Casas:2001sr} for encoding the  seesaw parameters. 
In the basis where the heavy neutrino singlets, the charged-lepton Yukawa matrix, and gauge 
interactions all appear flavour-diagonal, and where $\Yn$ denotes the seesaw Yukawa couplings, 
the product $\Yn^\dagger\Yn$ can be written as,
\be
\label{CI_par}
\Yn^\dagger\Yn = U_{\mathrm{PMNS}}\sqrt{\kappa} R^\dagger \mathcal{M} R \sqrt{\kappa}U_{\mathrm{PMNS}}^\dagger ,
\ee
where $\kappa$ contains the light neutrino masses inferred from low-energy experiments,
\bea
\kappa &=& \frac{\mathcal{M}_\nu}{\langle H^0_2 \rangle^2} \nonumber \\
\mathcal{M}_\nu &=& \mathrm{diag}\left(m_{\nu_1},m_{\nu_2},m_{\nu_3}\right) ,
\eea
and $\mathcal{M}$ denotes the diagonal Majorana singlet mass matrix, 
$\mathrm{diag}\left(\mathcal{M}_1,\mathcal{M}_2,\mathcal{M}_3\right)$. 
The matrix $U_{\mathrm{PMNS}}$ labels the neutrino mixing matrix inferred from the neutrino 
oscillation data, and the orthogonal matrix $R$ contains, in principle, three additional complex 
mixing parameters originating at the Majorana mass scale. Eq.(\ref{CI_par}) provides a general 
description of the seesaw mechanism and gives a useful parametrization for examining 
lepton-flavour violation (LFV) in seesaw models.

In order to determine the level of LFV in a given model, the full MSSM RGEs must be integrated 
from the unification scale to the weak scale, integrating each Majorana gauge-singlet neutrino 
out successively at its appropriate mass scale.
In the following sections we examine two interesting limiting cases of the seesaw: strongly hierarchical Majorana gauge-singlet neutrinos and degenerate Majorana 
gauge-singlet neutrinos, both with hierarchical light neutrinos, and each with the Majorana scale set at 
$M_R = 10^{14}$ GeV. This restriction reduces the number of free parameters in the orthogonal 
matrix $R$. 

\section{Hidden-Sector Renormalization Effects on LFV at Leading-Logarithmic Order}
\label{leading_log}

As a demonstration of the basic idea, we consider the toy self-interacting hidden-sector presented 
in~\cite{Cohen:2006qc,Dine:2004dv}, which contains the cubic superpotential term
\be
\label{Inter}
W_{\mathrm{h}} = \frac{\lambda}{3!} X^3 .
\ee
This simple superpotential cannot by itself break supersymmetry, and hence the hidden sector must 
contain additional interactions responsible for generating an F or D-term vacuum expectation value
(VEV) in any realistic model. For the purposes of examining the effects of hidden-sector 
renormalization, in this section we will suppose that eq.(\ref{Inter}) appears as the dominant 
self-interaction term in the hidden-sector superpotential, and that it provides the dominant 
hidden-sector contribution to the anomalous dimension of the operator mediating 
supersymmetry-breaking scalar masses in the observable sector. We assume that
the hidden-sector chiral superfields couple to the visible sector fields through the 
non-renormalizable operators
\be
\label{non_ren}
\int d^4\theta\hspace{.7mm} k_i \frac{X^\dagger X}{M_{\mathrm{Pl}}^2} \Phi^\dagger_i \Phi_i + \int d^2\theta\hspace{0.7mm} \omega \frac{X}{M_{\mathrm{Pl}}} W_n W_n,
\ee
where $\Phi_i$ and $W_n$ denote the MSSM chiral superfields and gauge fields respectively, and 
$M_{\mathrm{Pl}}$ denotes the scale of gravitational mediation -- the reduced Planck mass. 
Once supersymmetry breaks in the hidden sector, these terms yield the usual soft 
supersymmetry-breaking masses in the MSSM. In particular, $k_i$ generates the soft scalar mass 
terms, whilst $\omega$ determines the gaugino masses. In holomorphic renormalization schemes, 
non-renormalization theorems protect $\omega$ at all scales. Finally, the usual MSSM gauge and 
Yukawa interactions generate the MSSM RGEs. 

\begin{figure}[ht!]
   \newlength{\picwidthzz}
   \setlength{\picwidthzz}{2in}
   \begin{center}
	   \subfigure[][]{\resizebox{\picwidthzz}{!}{\includegraphics{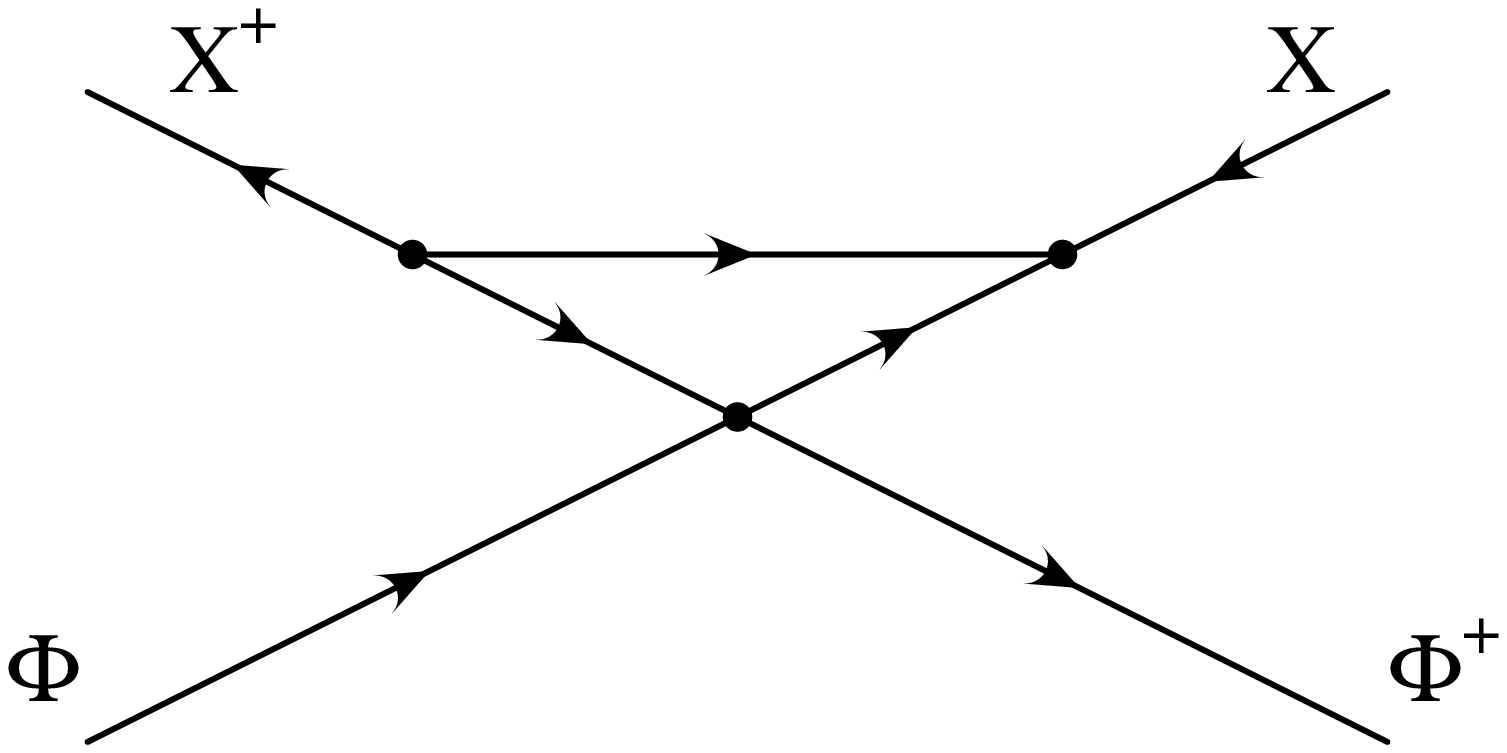}}}
	   \subfigure[][]{\resizebox{\picwidthzz}{!}{\includegraphics{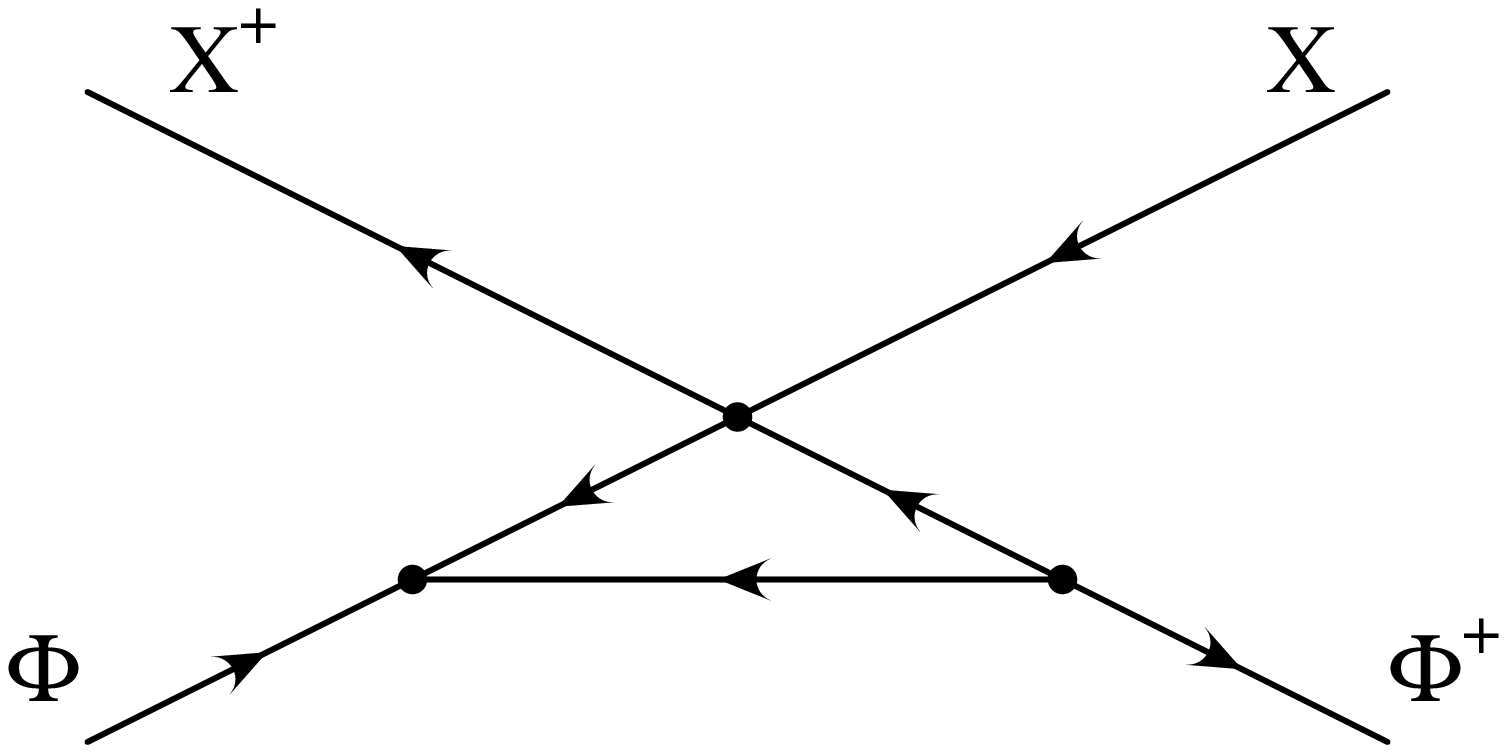}}}
       	   \subfigure[][]{\resizebox{\picwidthzz}{!}{\includegraphics{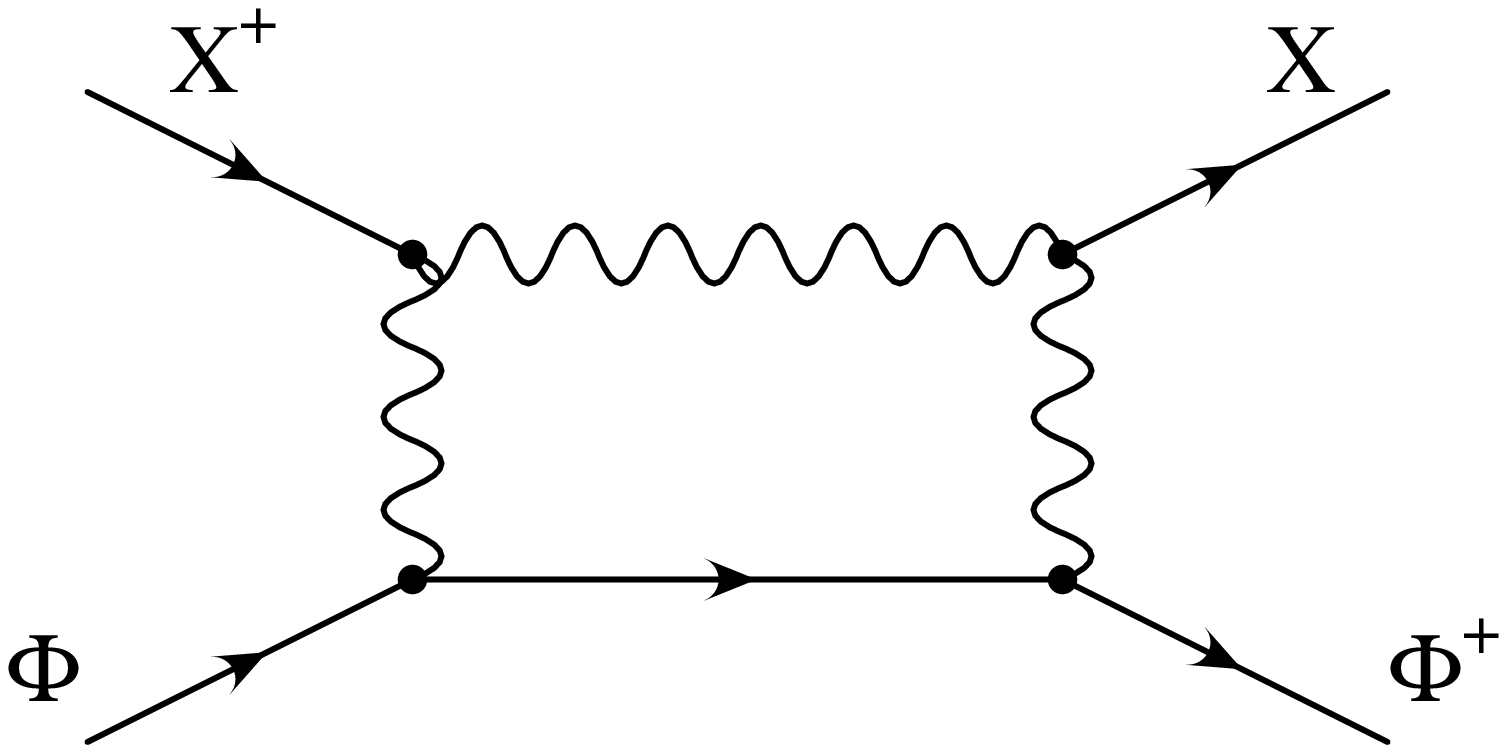}}}
   \end{center}
   \caption{\it One-loop supergraphs contributing to the scalar-mass renormalization. The first 
   diagram represents the quantum corrections arising from the hidden sector, whilst the last two 
   diagrams represent the usual visible-sector loop diagrams.}
   \label{super_graph}
\end{figure}

The coefficient $k_i$ in eq.(\ref{non_ren}), which we take to be flavour-diagonal at the unification scale, 
is renormalized by the two separate contributions given in Fig.~\ref{super_graph}: 
the usual visible-sector interactions involving vector and chiral superfields of the MSSM, and the 
hidden-sector self-interactions in eq.(\ref{Inter}). As a result, the RGE for $k_i$ in the presence of the 
hidden cubic superpotential reads
\be
\frac{d k_i}{dt} = \frac{2\lambda^*\lambda}{16\pi^2} k_i + \mathrm{gauge}\hspace{1mm}\mathrm{and}\hspace{1mm}\mathrm{visible}\hspace{1mm}\mathrm{sector}\hspace{1mm}\mathrm{contributions}.
\label{an_def}
\ee
The hidden-sector renormalization implies that the scalar-mass RGEs are augmented between the 
hidden-sector and messenger scales. In the hidden-sector theory under consideration, the RGE for 
an MSSM scalar sparticle becomes
\be
\frac{d{\bf m_S}^2}{dt} \rightarrow \frac{d{\bf m_S}^2}{dt} + \frac{2\lambda^*\lambda}{16\pi^2}{\bf m_S}^2.
\label{auger}
\ee
In general, the additional terms augmenting the usual MSSM RGEs arising from the hidden sector 
will be more complicated than the prescription given in eq.(\ref{auger}). To give a full description of 
the hidden-sector effect on the MSSM RGEs, a complete model of the hidden sector would be 
required~\cite{Murayama:2007ge,Dine:2004dv,Schmaltz:2006qs}. We stress that in this section we 
are simply considering a toy example that illustrates the effects on the level of slepton mixing, and 
hence of charged LFV, in supersymmetric models. We return to the problem of model dependence in following sections.

While eq.(\ref{auger}) demonstrates that we may expect shifts in the mass eigenvalues in the scalar 
sector of the MSSM, the hidden sector can also shift indirectly the off-diagonal elements. Given that the 
off-diagonal scalar mass-matrix elements determine the level of flavour-changing neutral currents in 
the MSSM, it is worth examining the size of this indirect effect. In particular, given that the seesaw sector 
sits near the hidden-sector scale in the model class we consider, the effect on the off-diagonal 
structure arising from the hidden sector will impact the amount of predicted charged LFV. To get an 
idea of the underlying physics of the effect, we proceed with our toy analytical example with scalar 
masses-squared running augmented by hidden-sector effects, but for analytic simplicity and clarity we ignore for the moment the running of the hidden-sector coupling $\lambda$ itself,
which is $O(\lambda^3)$.

In the MSSM with Majorana gauge-singlet neutrinos, the RGE of the left-handed slepton 
mass-squared reads (before including the hidden-sector contribution):
\bea
\label{slepton}
(16\pi^2)\frac{d{\bf m_L}^2}{dt} &=& {\bf m_L}^2 \Ye^\dagger\Ye + \Ye^\dagger \Ye {\bf m_L}^2 + {\bf m_L}^2 \Yn^\dagger\Yn + \Yn^\dagger \Yn {\bf m_L}^2 \nonumber \\
&& + 2\left(\Ye^\dagger {\bf m_e}^2 \Ye + m^2_{H_d} \Ye^\dagger\Ye + \Ae^\dagger \Ae\right) + 2\left(\Yn^\dagger {\bf m_\nu}^2 \Yn + m^2_{H_u} \Yn^\dagger\Yn + \An^\dagger \An\right) \nonumber \\
&&- \left(2 g_1^2 |M_1|^2 + 6 g_2^2 |M_2|^2\right){\bf I}_3 ,
\eea
where $t$ denotes the logarithm of the running scale, and $\Yn$ denotes the seesaw Yukawa 
couplings. In order to examine the leading-logarithmic behaviour for the prediction of LFV
in the presence of hidden-sector effects in this model class, we consider the following 
simplified 2-by-2 model case:
\be
\label{Simp}
\frac{d}{dt} \left[ \begin{array}{cc} y_{11} & y_{12} \\ y_{21} & y_{22} \end{array} \right] = \frac{1}{8\pi^2}\left(\lambda^2 \left[\begin{array}{cc} 1 & 0 \\ 0 & 1 \end{array}\right] \left[ \begin{array}{cc} y_{11} & y_{12} \\ y_{21} & y_{22} \end{array} \right] + 3\left[ \begin{array}{cc} n_{11} & n_{12} \\ n_{21} & n_{22} \end{array} \right] \left[ \begin{array}{cc} y_{11} & y_{12} \\ y_{21} & y_{22} \end{array} \right] \right) .
\ee
In eq.(\ref{Simp}), we model the hidden-sector effect by the term proportional to $\lambda$. 
In this case, the matrix
\be
\left[ \begin{array}{cc} y_{11} & y_{12} \\ y_{21} & y_{22} \end{array} \right]
\ee
is analogous to the slepton mass-squared matrix, and
\be
\left[ \begin{array}{cc} n_{11} & n_{12} \\ n_{21} & n_{22} \end{array} \right]
\ee
is analogous to the matrix product $\Yn^\dagger\Yn$. We place a factor of three in eq.(\ref{Simp}) to 
match the factor three that appears in the leading-logarithmic analysis of eq.(\ref{slepton}). In our approximation we ignore the possible presence of A-terms. The initial conditions at the unification 
scale read:
\be
\left.\left[ \begin{array}{cc} y_{11} & y_{12} \\ y_{21} & y_{22} \end{array} \right]\right|_0 = \left[\begin{array}{cc} m_0^2 & 0 \\ 0 & m_0^2 \end{array}\right] ,
\ee
and we assume that $y_{11}, y_{22} \gg y_{12}, y_{21}$ throughout. Note that the element $y_{21}$ 
controls the level for the branching ratio $\meg$. Using the above matrix differential equation, we 
obtain a coupled set of differential equations, namely,
\bea
\frac{d y_{21}}{dt} &=& \frac{3}{8\pi^2}\left(n_{22} y_{21} + n_{21} y_{11}\right) \\
\frac{d y_{11}}{dt} &=& \frac{1}{8\pi^2}\left( (\lambda^2 + 3n_{11}) y_{11} + 3n_{12} y_{21}\right).
\eea
In the limit that $y_{11} \gg y_{21}$, the approximate solution for $y_{11}$ becomes
\be
y_{11} \approx m_0^2 e^{(\lambda^2 + 3n_{11})/8\pi^2 t} 
\ee
and using this solution, we obtain
\bea
\frac{d y_{21}}{dt} &=& \frac{3}{8\pi^2}\left(n_{22} y_{21} + n_{21} m_0^2 e^{(\lambda^2 + 3n_{11})/8\pi^2 t}\right) \nonumber \\
&\approx& \frac{3n_{21} m_0^2}{8\pi^2} e^{(\lambda^2 + 3n_{11})/8\pi^2 t} ,
\eea
where the approximation again makes use of $y_{11} \gg y_{21}$. Accordingly, the approximate 
solution for $y_{21}$ reads,
\be
y_{21} = \frac{3n_{21} m_0^2}{\lambda^2 + 3n_{11}} \left( e^{(\lambda^2 + 3n_{11})/8\pi^2 t} - 1 \right).
\ee
Integrating the renormalization group flow to the seesaw scale yields 
$t = -\ln(M_{\mathrm{X}}/M_{\mathrm{R}})$, which allows us to write
\be
y_{21} = \frac{3n_{21} m_0^2}{\lambda^2 + 3n_{11}} \left( e^{-(\lambda^2 + 3n_{11})/8\pi^2 \ln(M_{\mathrm{X}}/M_{\mathrm{R}})} - 1 \right).
\label{master_lfv}
\ee
Two limiting cases of eq.(\ref{master_lfv}) are worth exploring:
\begin{itemize}
\item{$\lambda^2 \ll 1$}
In this limit, we recover the usual leading-logarithmic result for seesaw-induced off-diagonal slepton 
mass terms, namely
\bea
y_{21} &\approx& - \frac{3n_{21} m_0^2}{8\pi^2} \ln\left(\frac{M_{\mathrm{X}}} {M_{\mathrm{R}}}\right) \nonumber \\
&=& - \frac{3\left(\Yn^\dagger\Yn\right)_{21} m_0^2}{8\pi^2} \ln\left(\frac{M_{\mathrm{X}}} {M_{\mathrm{R}}}\right) ,
\eea
where we have made the identification $n_{21} \rightarrow \left(\Yn^\dagger\Yn\right)_{21}$ in the last 
line above.

\item{$\lambda^2 \ln(M_{\mathrm{X}}/M_{\mathrm{R}}) \gtrsim 8\pi^2$}
In this limit, assuming $\lambda \gg n_{11}$, we now have 
\bea
y_{21} &\approx& \frac{3n_{21} m_0^2}{\lambda^2}\left(\left( \frac{M_{\mathrm{X}}}{M_{\mathrm{R}}} \right)^{-\lambda^2/8\pi^2} -1\right) \nonumber \\
&\approx& \frac{3\left(\Yn^\dagger\Yn\right)_{21} m_0^2}{\lambda^2}\left(\left( \frac{M_{\mathrm{X}}}{M_{\mathrm{R}}} \right)^{-\lambda^2/8\pi^2} -1\right) ,
\eea
where again we made the identification $n_{21} \rightarrow \left(\Yn^\dagger\Yn\right)_{21}$. 
We see that the $y_{21}$ depends on $1/\lambda^2$, which can lead to a suppression in the 
mixing for large $\lambda$.
\end{itemize}

\begin{figure}[ht!]
   \newlength{\picwidthaa}
   \setlength{\picwidthaa}{2.7in}
   \begin{center}
       \resizebox{\picwidthaa}{!}{\includegraphics{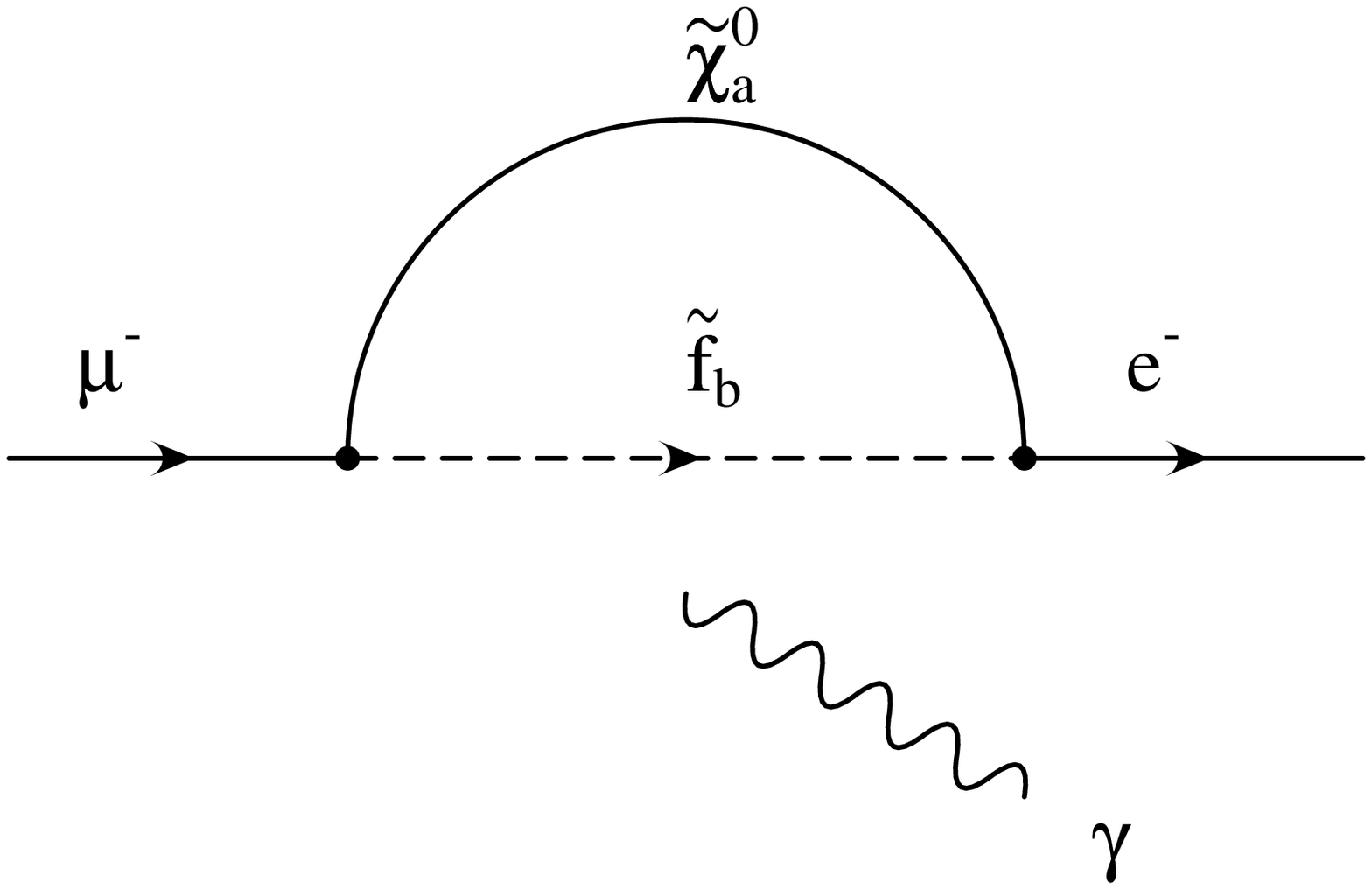}}
       \resizebox{\picwidthaa}{!}{\includegraphics{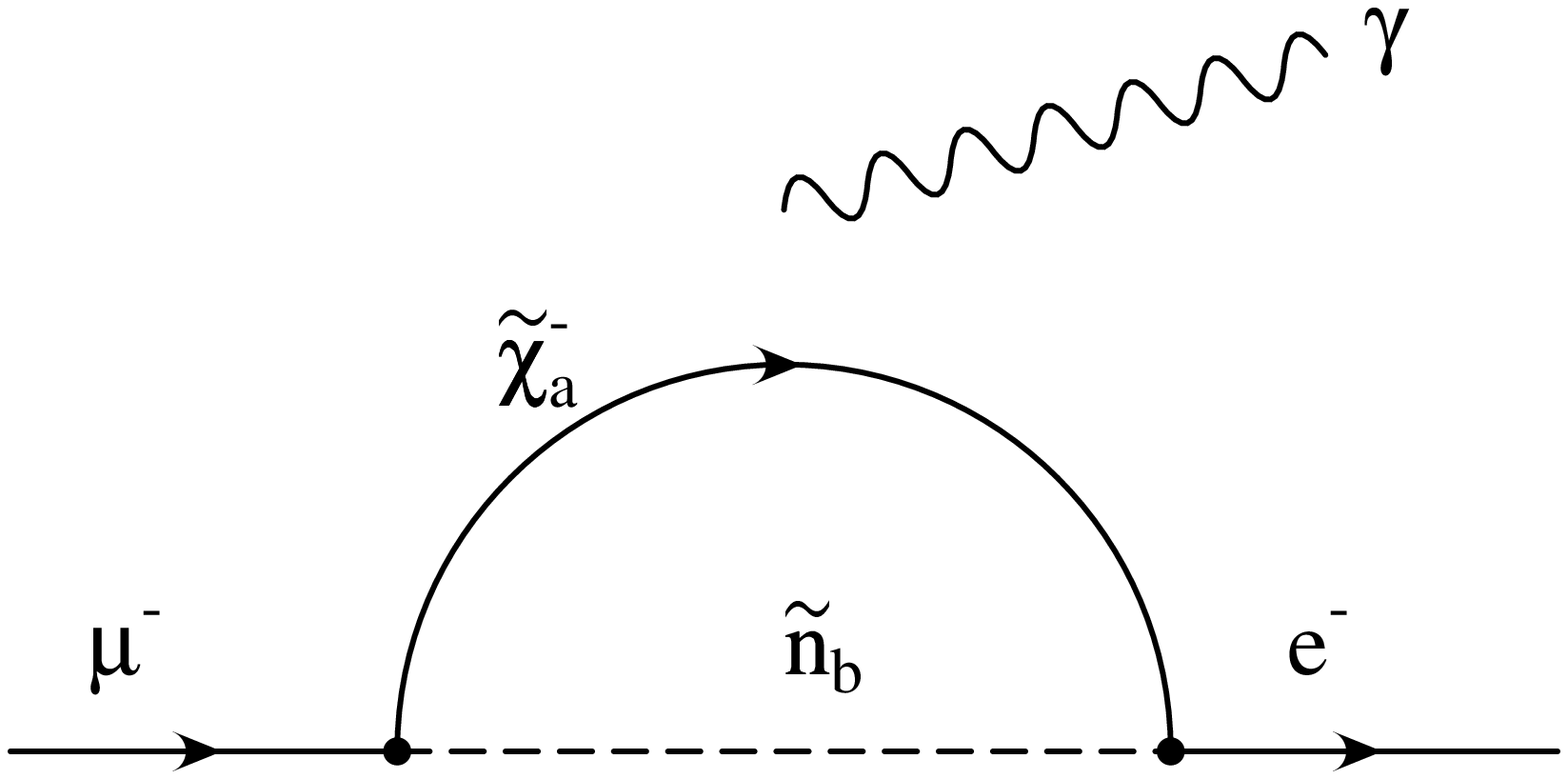}}
   \end{center}
   \caption{\it One-loop diagrams contributing to $\meg$ in the MSSM. The symbols
   $\tilde f_b$ denote charged 
   sleptons, $\tilde n_b$ denote sneutrinos associated with the left-handed slepton doublet, 
   $\tilde\chi^\pm$ denote the charginos, and $\tilde\chi^0$ denote the neutralinos.}
   \label{loop_dia}
\end{figure}

The branching ratio for $\meg$ depends at leading order on the off-diagonal slepton mass matrix, 
${\bf m_L}_{21}$, via
\be
\mathrm{BR}(\meg) \approx \frac{\alpha^3}{G_F^2} \frac{\left|{\bf m_L}_{21}^2\right|^2}{m^8_S}\tan^2\beta ,
\ee
where $m_S$ denotes a typical sparticle mass. The leading-logarithmic approximation for the 
off-diagonal slepton mass matrix elements yields
\be
\label{slepton_L}
{\bf m_L}_{ij}^2 \approx -\frac{m_0^2}{8\pi^2} (3 + a_0^2) \ln\left(\frac{M_{\mathrm{X}}}{M_R}\right) (\Yn^\dagger\Yn)_{ij} ,
\ee
where $m_0^2$ denotes the universal scalar mass, $a_0$ labels the constant of proportionality 
in the A-terms, and $M_R$ denotes the Majorana mass scale. The branching ratio for 
$\meg$ therefore becomes
\be
\mathrm{BR}(\meg) \sim \frac{\alpha^3}{G_F^2m_S^8} \left|  -\frac{m_0^2}{8\pi^2} (3 + a^2) \ln\left(\frac{M_{\mathrm{X}}}{M_R}\right)\right|^2 \left|(\Yn^\dagger\Yn)_{21}\right|^2 \tan^2\beta.
\ee
The current experimental bound reads $\mathrm{BR}(\meg) < 1.2\times 10^{-11}$, and the 
MEG experiment at PSI expects to attain a sensitivity at a level of 
$\mathrm{BR}(\meg) \sim 5\times 10^{-14}$~\cite{Mori:2002sg}. In the next section we calculate 
the branching ratio for $\meg$ using the full one-loop expression arising from Fig.~\ref{loop_dia}, 
and we run the full one-loop RGEs for the MSSM, in classes of parametrizations of the hidden sector.
\begin{figure}[ht!]
   \newlength{\picwidtha}
   \setlength{\picwidtha}{5.5in}
   \begin{center}
       \resizebox{\picwidtha}{!}{\includegraphics{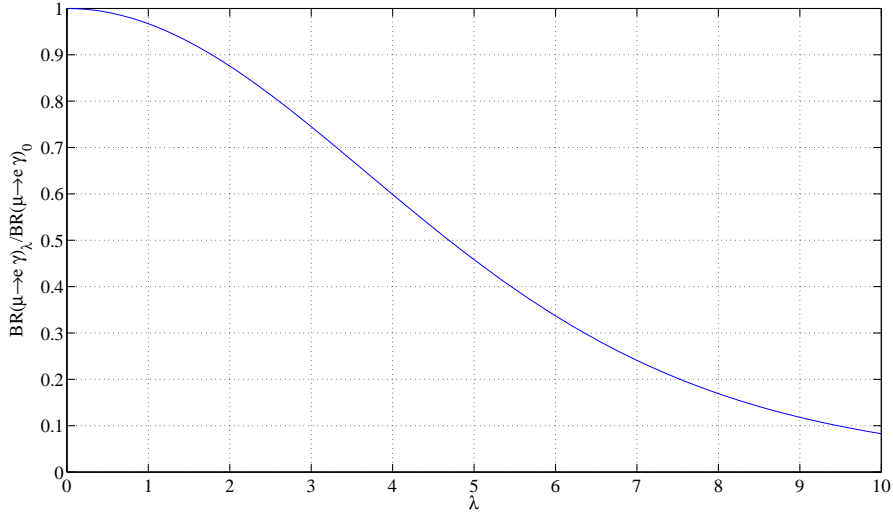}}
   \end{center}
   \caption{\it The branching ratio for $\mu \to e \gamma$ calculated incorporating the hidden sector in
   leading-logarithmic order as a function of 
   $\lambda$, relative to the value without hidden-sector effects.}
   \label{COMP}
\end{figure}

Using eq.(\ref{slepton_L}), the branching ratio for $\mu \rightarrow e \gamma$ depends on $y_{21}$ as
\be
\mathrm{BR}(\mu\rightarrow e \gamma) \sim \frac{\alpha^3}{G_F^2} \frac{|y_{21}|^2}{\tilde m_S^8} \tan^2\beta.
\label{BR_ratio_again}
\ee
We see from Fig.~\ref{COMP} that the suppression of the branching ratio as a function of the fixed 
hidden-sector coupling becomes larger than a factor of two when $\lambda \gtrsim 5$. We 
should emphasize that in the above analysis we did not consider the shift in the scalar spectrum 
relative to the MSSM from hidden-sector dynamics, nor did we include the renormalization-group 
running of $\lambda$. Since the branching ratio depends inversely on sparticle masses to the eighth 
power (see eq.(\ref{BR_ratio_again})), the changes in the sparticle spectra will have a non-negligible 
effect on the predicted rate. 
A competition will emerge between the suppression expected from the above analysis and the shifts in the sparticle spectra. We anticipate that in some regions where the hidden sector lowers the scalar spectrum 
sufficiently relative to the unaltered MSSM, the light scalars will dominate over the suppression factor, 
and yield an enhanced rate for $\meg$. We explore these details in the next section.

Our analysis thus far assumed $\lambda$ was constant over the range of integration, in order to 
allow simple analytic treatment. In reality, the hidden-sector coupling $\lambda$ runs with its own 
$\beta$ function: $\beta_{\lambda} = (3/(32\pi^{2}))\lambda^{3}$. 
However, since the superpotential of eq.(\ref{Inter}) by itself does not break supersymmetry, we must 
have additional hidden-sector interactions, so this model is just a toy, and not to be taken literally in 
detail. The point of the above discussion was simply to illustrate how a hidden sector self-interaction 
can influence the level of predicted charged LFV. In the following section, we examine the impact of 
the hidden sector in a model-independent fashion. Instead of hypothesizing a particular form of the 
coupling in the hidden sector, we parametrize the anomalous dimension itself. This parametrization 
allows us to examine both IR- and UV-free theories with ease.

\section{LFV with general Hidden-Sector Effects}

\begin{figure}[ht!]
   \newlength{\picwidthbb}
   \setlength{\picwidthbb}{7in}
   \begin{center}
       \resizebox{\picwidthbb}{!}{\includegraphics{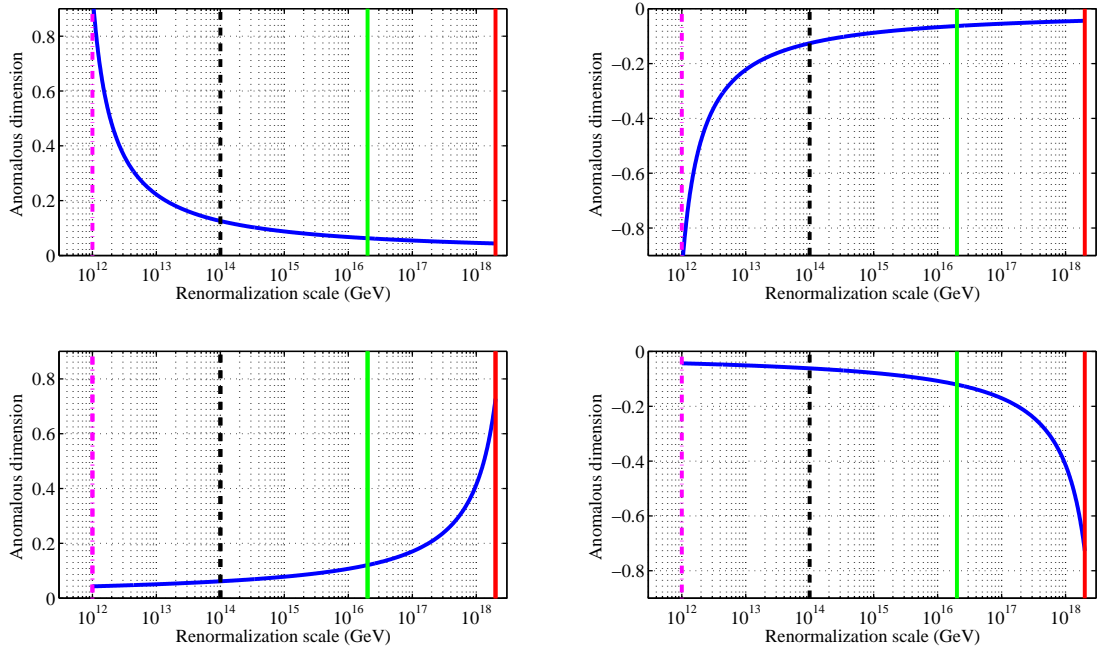}}
   \end{center}
   \caption{\it Four panels demonstrating the possible behaviours of the anomalous dimension 
   outlined in the text. Top left: $b_\gamma = 2/3$, $a_\gamma = \log(5\times 10^{11}\hspace{1mm} 
   \mathrm{GeV}/\mu)$, top right: $b_\gamma = -2/3$, $a = \log(5\times 10^{11}\hspace{1mm} 
   \mathrm{GeV}/\mu)$, bottom left: $b_\gamma = -2/3$, $a = \log(5\times 10^{18}\hspace{1mm} 
   \mathrm{GeV}/\mu)$, bottom right: $b_\gamma = 2/3$, $a = \log(5\times 10^{18}\hspace{1mm} 
   \mathrm{GeV}/\mu)$. In each panel, the heavy solid line at $2\times 10^{18}$ GeV denotes the 
   reduced Planck mass, the light solid line at $2\times 10^{16}$ GeV denotes the unification scale, 
   the heavy dashed line at $10^{14}$ GeV represents the Majorana mass scale, and the light 
   dashed-dotted line at $10^{12}$ GeV represents the hidden-sector mass scale.}
   \label{ano_dim}
\end{figure}

We saw from eq.(\ref{an_def}) that in the cubic hidden superpotential theory we had
\be
\frac{d k_i}{dt} = \frac{2\lambda^*\lambda}{16\pi^2} k_i + 
\mathrm{visible}\hspace{1mm}\mathrm{sector}\hspace{1mm}\mathrm{contributions}.
\ee
In order to examine general theories of the hidden sector, we require a parametrization that does 
not depend on the particulars of the hidden-sector implementation. For a general hidden sector 
we can define
\be
\label{an_redef}
\frac{d k_i}{dt} = \gamma(t) k_i + 
\mathrm{visible}\hspace{1mm}\mathrm{sector}\hspace{1mm}\mathrm{contributions} ,
\ee
where $\gamma(t)$ denotes the anomalous dimension contributed by the hidden sector. Instead of 
using $\gamma(t)$ from some specific theory as we did in the previous section, following our 
previous paper~\cite{CEM1}, we consider the parametrization~\cite{Ross},
\be
\gamma(t) = \frac{1}{b_\gamma(s-a_\gamma)} ,
\ee
where $b_\gamma$ and $a_\gamma$ are theory-dependent factors, and $s$ denotes the logarithm
of the running scale. Depending on the location of the pole in $\gamma(t)$, we can examine both IR-
and UV-free theories. We consider the following four cases for the anomalous dimension 
$\gamma(t)$: $b_\gamma = \pm 2/3$, $a_\gamma = \log(5\times 10^{11} \mathrm{GeV}/\mu),
 \log(5\times 10^{18} \mathrm{GeV}/\mu)$. We depict these cases in Fig.~\ref{ano_dim}.

By choosing the poles in either of two locations -- just beyond the reduced Planck mass, or just below 
the hidden-sector scale ($10^{12}$ GeV) -- we arrange that the hidden sector remains perturbative over 
the integration range and up to the reduced Planck mass itself. Once we reach the hidden-sector 
mass scale, we integrate out the hidden-sector physics, returning to the usual MSSM RGEs. 
As we see in Fig.~\ref{ano_dim}, by placing the pole above the reduced Planck mass in the IR-free 
case, the magnitude of the anomalous dimension in the interval between the unification and the 
Majorana mass scales is smaller than in our UV-free case. We make this choice in order to ensure 
that the Landau pole in the IR-free case does not appear below the reduced Planck mass. In the 
following section, we will apply each of these cases to the MSSM with the Majorana scale set at 
$10^{14}$~GeV and with both hierarchical and degenerate Majorana gauge-singlet neutrinos, under 
the assumption of a normal hierarchy for the light neutrinos of the Standard Model. To determine the 
effect of the hidden sector on the predicted rate for $\meg$, we run numerically the seesaw-extended 
MSSM RGEs including the hidden sector from the unification scale to the weak scale. 
We integrate out the Majorana gauge-singlet neutrinos at their associated scale and we also integrate 
out the hidden-sector effect at the hidden sector scale of $10^{12}$~GeV.  

\subsection{Hierarchical Gauge-Singlet Neutrinos}

We recall from section \ref{LFV_MSSM} that the branching ratio $BR(\meg)$ depends on the 
combination $\Yn^\dagger\Yn$, which through the Casas-Ibarra parametrization~\cite{Casas:2001sr} 
can be expressed in terms of an orthogonal matrix $R$. In the case of hierarchical gauge-singlet 
Majorana neutrinos, one finds that (in the notation of section {\ref{LFV_MSSM}),
\be
(\Yn)_{ij} \approx \sqrt{\mathcal{M}_3}\delta_{i3} R_{3l}(\sqrt{\kappa})_l U^\dagger_{lj}.
\ee
Thus, the branching ratio for $\meg$ depends almost exclusively on one mixing angle in $R$, 
which we denote as $\theta_1$ (we are working under the assumption that $R$ is real). To 
demonstrate the effect of the hidden sector on the branching ratio, we choose a typical point in the 
MSSM parameter space: $\tan\beta = 20$, $\mu >0$, $a_0 = 0$, $M_0 = 200$ GeV, 
$M_{1/2} = 760$ GeV. In all cases, we set the hidden-sector mass scale at $10^{12}$ GeV. 

We show in Fig.~\ref{Four_panel} the effect of the hidden sector on the branching ratio for $\meg$ 
as a function of $\theta_1$. In each panel the solid blue curve denotes the prediction for the branching 
ratio in the absence of hidden-sector self-interactions, and the dashed red curve represents the 
prediction with the hidden-sector effect turned on. The top left panel displays the result for 
$b_\gamma = 2/3$, $a_\gamma = \log(5\times 10^{11}\hspace{1mm} \mathrm{GeV}/\mu)$,
the top right panel displays the result for $b_\gamma = -2/3$, $a = \log(5\times 10^{11}
\hspace{1mm} \mathrm{GeV}/\mu)$, the bottom left panel displays the result for 
$b_\gamma = 2/3$, $a = \log(5\times 10^{18}\hspace{1mm} \mathrm{GeV}/\mu)$, and the bottom 
left panel displays the result for $b_\gamma = -2/3$, $a = \log(5\times 10^{18}\hspace{1mm} 
\mathrm{GeV}/\mu)$.
 
\begin{figure}[ht!]
   \newlength{\picwidthb}
   \setlength{\picwidthb}{7in}
   \begin{center}
       \resizebox{\picwidthb}{!}{\includegraphics{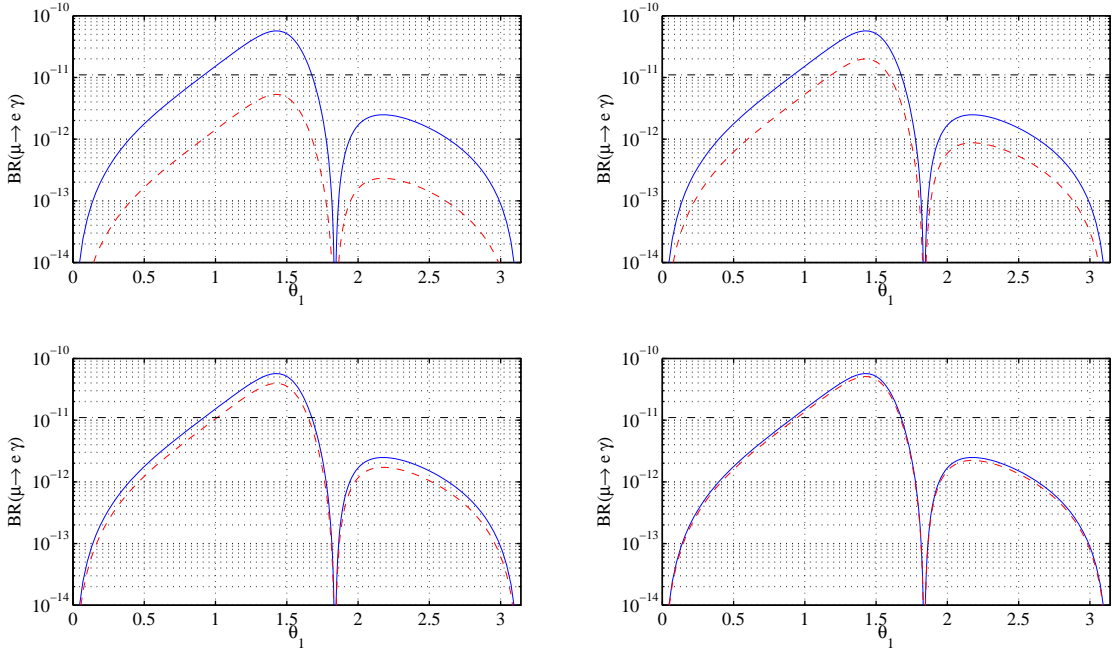}}
   \end{center}
   \caption{\it Four panels demonstrating the effect on $\meg$ for the four behaviours of the 
   anomalous dimension outlined in the text. Top left: $b_\gamma = 2/3$, $a_\gamma = 
   \log(5\times 10^{11}\hspace{1mm} \mathrm{GeV}/\mu)$, top right: $b_\gamma = -2/3$, 
   $a = \log(5\times 10^{11}\hspace{1mm} \mathrm{GeV}/\mu)$, bottom left: $b_\gamma = 2/3$, 
   $a = \log(5\times 10^{18}\hspace{1mm} \mathrm{GeV}/\mu)$, and bottom right: $b_\gamma = -2/3$, 
   $a = \log(5\times 10^{18}\hspace{1mm} \mathrm{GeV}/\mu)$.}
   \label{Four_panel}
\end{figure}

We see that the largest suppression of the rate occurs for the UV-free anomalous dimension 
displayed in the top two panels. This observation simply reflects that in the UV-free setting we 
chose the pole of the anomalous dimension at $5\times 10^{11}$ GeV, a factor of two lower than 
the hidden sector scale, whilst in the IR-free case we placed the pole above the reduced Planck 
mass. As a result of these choices, the UV-free case has a larger anomalous dimension over the 
integration range. In this UV-free case, the anomalous dimension grows during the running of 
the RGEs from the unification scale to the hidden sector scale. Since the seesaw remains active 
until the Majorana scale is reached, we see that the growth in the hidden-sector anomalous 
dimension in the UV-free scenario serves to influence increasingly the rate for $\meg$ -- in this 
case suppressing the rate by up to an order of magnitude. This observation is consistent with our 
semi-analytic treatment in the previous section. 

We can see how the allowed parameter space changes with the hidden-sector physics by 
examining the allowed parameter regions in the conventional CMSSM $M_0-M_{1/2}$ plane. 
In Fig.~\ref{Hi_UV_free} we examine the hierarchical Majorana gauge-singlet neutrino case with 
$\theta_1 = 1.4$, displaying the large effect that the hidden sector may have 
on the predicted rate for $\meg$. The solid white area in each of the three panels indicates
the region that is currently excluded by the experimental upper limit ($\mathrm{BR}(\meg) < 
1.2\times 10^{-11}$). The left panel displays the allowed region with the hidden sector effect turned off; 
the second panel shows the effect with the allowed regions for $b_\gamma = 2/3$, 
$a_\gamma = \log(5\times 10^{11}\hspace{1mm} \mathrm{GeV}/\mu)$, and the third panel 
displays the effect with $b_\gamma = -2/3$, $a_\gamma = \log(5\times 10^{11}\hspace{1mm} 
\mathrm{GeV}/\mu)$. In each case we set the Majorana mass scale to $10^{14}$~GeV and the 
hidden-sector scale to $10^{12}$~GeV. 

\begin{figure}[ht!]
   \newlength{\picwidthc}
   \setlength{\picwidthc}{7in}
   \begin{center}
       \resizebox{\picwidthc}{!}{\includegraphics{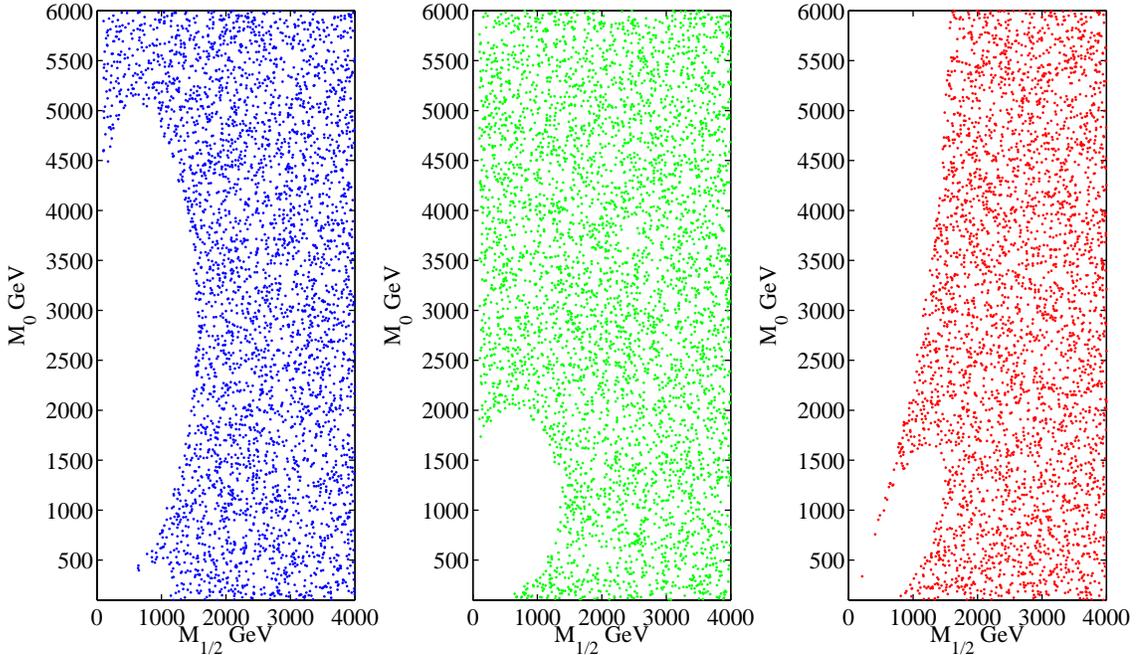}}
   \end{center}
   \caption{\it The allowed parameter space with $BR(\meg) < 1.2\times 10^{-11}$ in the
   presence of hierarchical gauge-singlet neutrinos and a UV-free hidden-sector theory, with
   the solid white regions denoting excluded regions. Left panel: allowed parameter space without 
   hidden-sector dynamics, middle panel: allowed parameter region with $b_\gamma = 2/3$, 
   $a_\gamma = \log(5\times 10^{11}\hspace{1mm} \mathrm{GeV}/\mu)$, and right panel: 
   allowed parameter region with $b_\gamma = -2/3$, $a_\gamma = \log(5\times 10^{11}
   \hspace{1mm} \mathrm{GeV}/\mu)$. In all cases $M_R = 10^{14}$~GeV, $\mu >0$, 
   $\tan\beta = 20$, $a_0=0$.}
   \label{Hi_UV_free}
\end{figure}

We see that the hidden sector can change the allowed 
parameter space quite dramatically. Comparing the last two panels with the first one, we see that 
some regions that were excluded have become allowed and regions that were allowed have become excluded. These figures indicate the subtle effects the hidden sector may have. From the 
semi-analytic treatment in the previous section, we saw that for a large enough hidden-sector 
coupling, we expect the rate for $\meg$ to become suppressed. However, as mentioned earlier,
the hidden sector also alters the sparticle spectrum, specifically the mass eigenvalues of the scalar 
particles, and this has a nontrivial effect on the branching ratio for $\meg$. In order to determine 
the full effect of the hidden sector on the branching ratio for $\meg$, we require a full numerical 
treatment as displayed in Fig.~\ref{Hi_UV_free}. In regions where the hidden sector serves to 
eliminate parameter space, the change in the sparticle spectrum is the dominant effect.

\begin{figure}[ht!]
   \newlength{\picwidthd}
   \setlength{\picwidthd}{7in}
   \begin{center}
       \resizebox{\picwidthd}{!}{\includegraphics{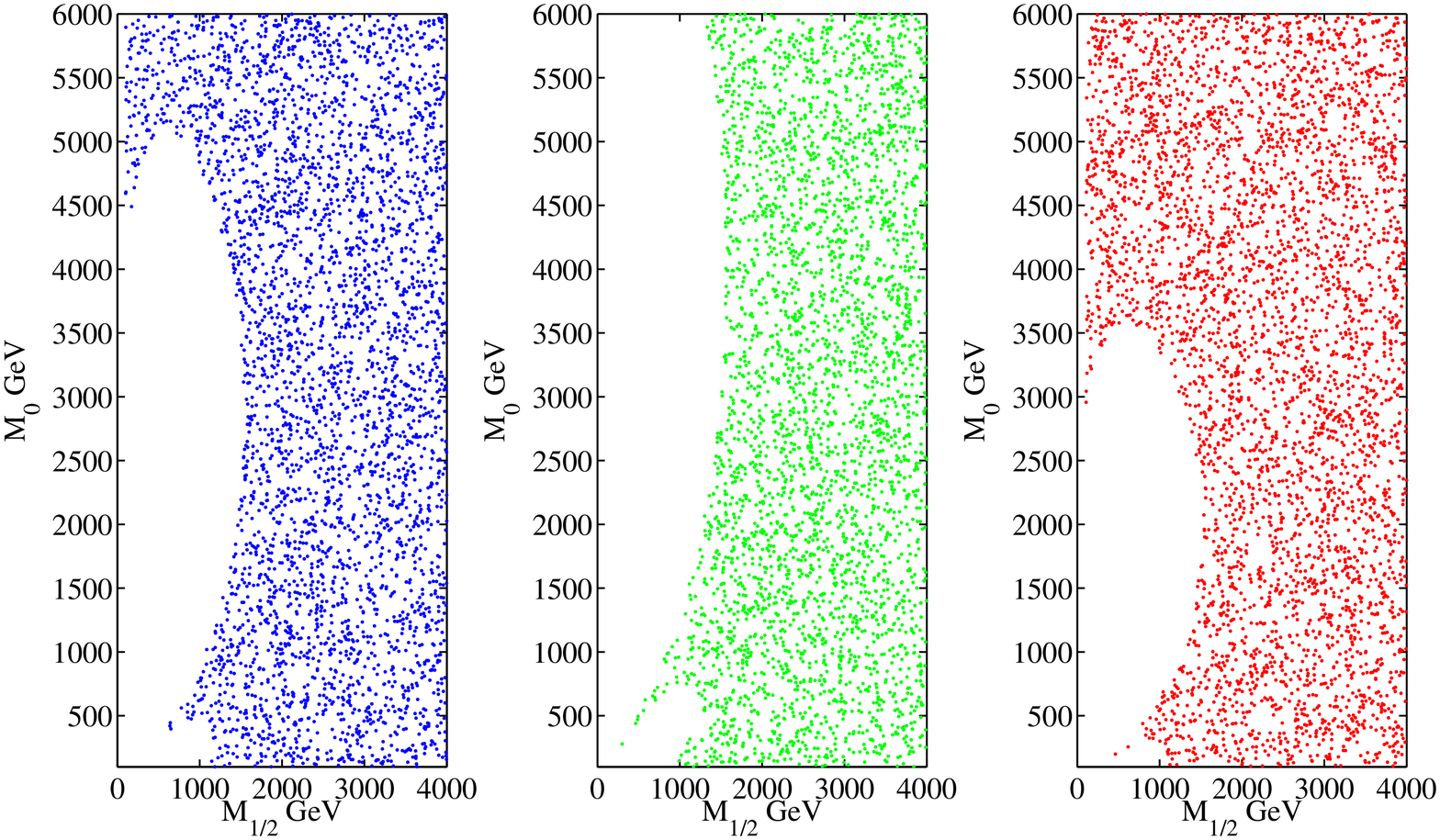}}
   \end{center}
   \caption{\it Hierarchical gauge-singlet neutrinos with an IR-free hidden-sector theory,
   displaying the allowed parameter space with $BR(\meg) < 1.2\times 10^{-11}$, with the solid white 
 regions denoting excluded regions. Left panel: allowed parameter space without hidden sector 
 dynamics, middle panel: allowed parameter region with $b_\gamma = 2/3$, $a_\gamma = 
 \log(5\times 10^{18}\hspace{1mm} \mathrm{GeV}/\mu)$, and right panel: allowed parameter region 
 with $b_\gamma = -2/3$, $a_\gamma = \log(5\times 10^{18}\hspace{1mm} \mathrm{GeV}/\mu)$. 
 In all cases $M_R = 10^{14}$ GeV, $M_{\mathrm{hidden}}\sim 10^{12}$, $\mu >0$, $\tan\beta = 20$,
 and $a_0=0$.}
   \label{Hi_IR_free}
\end{figure}

In Fig.~{\ref{Hi_IR_free}, we show the effect of the hidden sector with an IR-free anomalous 
dimension in the same parameter space as Fig.~{\ref{Hi_UV_free}. Again, we see that the hidden 
sector can change dramatically the allowed parameter space. The first panel displays the allowed 
region with the hidden-sector effect turned off, the second panel shows the effect on the allowed 
regions for $b_\gamma = 2/3$, $a_\gamma = \log(5\times 10^{18}\hspace{1mm} \mathrm{GeV}/\mu)$,
and the third panel displays the effect with $b_\gamma = -2/3$, $a_\gamma = \log(5\times 10^{18}
\hspace{1mm} \mathrm{GeV}/\mu)$. In all cases we have taken $\theta_1 =1.4$ with the Majorana 
mass scale at $10^{14}$~GeV and the hidden sector mass scale at $10^{12}$~GeV. 

\subsection{Degenerate Gauge-Singlet Neutrinos}

In the degenerate gauge-singlet neutrino case the combination $\Yn\Yn^\dagger$  yields no 
dependence on any particular angle in the orthogonal matrix $R$. In this case, once 
$\Yn\Yn^\dagger$ is given, the branching ratio is fully determined. We begin by displaying the 
parameter space in the $M_0-M_{1/2}$ plane. 
The three panels in Fig.~\ref{D_UV_free} show the changes in the allowed parmeter space in the 
presence of the hidden sector. The first panel displays the allowed region with the hidden sector 
effect turned off, the second panel shows the effect with the allowed regions for $b_\gamma = 2/3$, 
$a_\gamma = \log(5\times 10^{11}\hspace{1mm} \mathrm{GeV}/\mu)$, and the third panel displays 
the effect with $b_\gamma = -2/3$, $a_\gamma = \log(5\times 10^{11}\hspace{1mm} \mathrm{GeV}/\mu)$. 
In all cases we have taken $\theta_1 =1.4$ with the Majorana scale at $10^{14}$ GeV and the 
hidden-sector scale at $10^{12}$ GeV. 

\begin{figure}[ht!]
   \newlength{\picwidthe}
   \setlength{\picwidthe}{7in}
   \begin{center}
       \resizebox{\picwidthe}{!}{\includegraphics{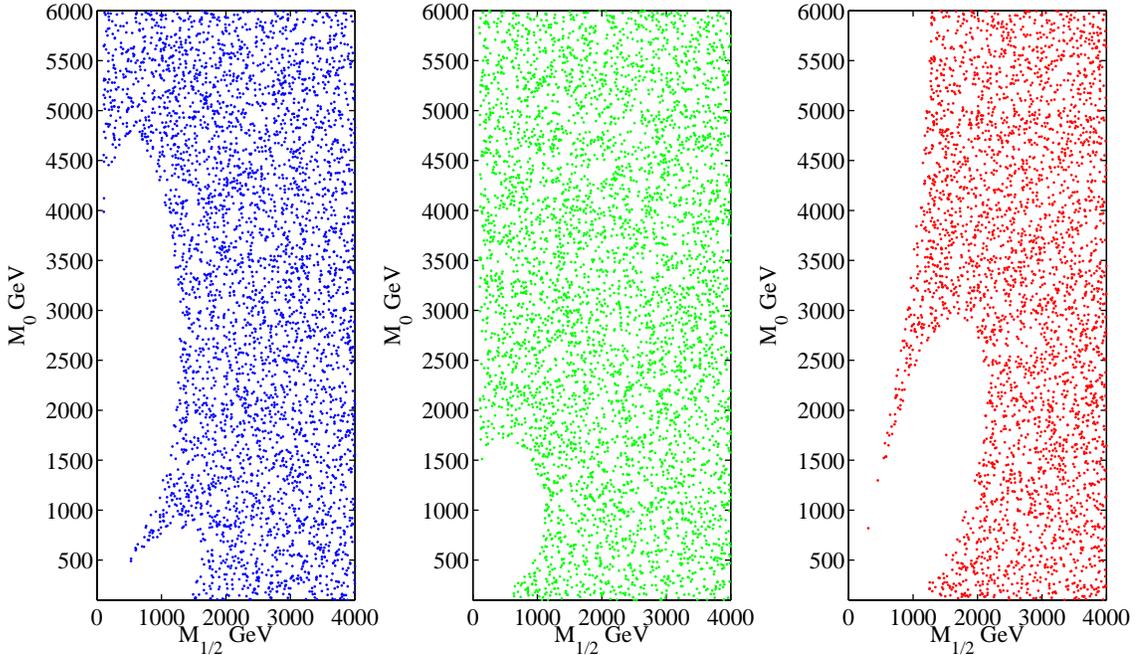}}
   \end{center}
   \caption{\it The case of degenerate gauge-singlet neutrinos with an UV-free hidden-sector theory,
   displaying the allowed parameter space with $BR(\meg) < 1.2\times 10^{-11}$, with the solid white
   regions denoting excluded regions. Left panel: allowed parameter space without hidden-sector 
   dynamics, middle panel: allowed parameter region with $b_\gamma = 2/3$, $a_\gamma = 
   \log(5\times 10^{11}\hspace{1mm} \mathrm{GeV}/\mu)$, and right panel: allowed parameter region 
   with $b_\gamma = -2/3$, $a_\gamma = \log(5\times 10^{11}\hspace{1mm} \mathrm{GeV}/\mu)$. 
   In all cases $M_R = 10^{14}$~GeV, $M_{\mathrm{hidden}}=10^{12}$~GeV, $\mu >0$, 
   $\tan\beta = 20$, and $a_0=0$.}
   \label{D_UV_free}
\end{figure}

In Fig.~\ref{D_IR_free} we repeat the analysis for the IR-free anomalous dimension case with 
parameters  $b_\gamma = 2/3$, $a_\gamma = \log(5\times 10^{18}\hspace{1mm} \mathrm{GeV}/\mu)$.
The third panel displays the effect with $b_\gamma = -2/3$, $a_\gamma = \log(5\times 10^{18}\hspace{1mm} \mathrm{GeV}/\mu)$, respectively. Again we see the competition between the expected 
suppression in the rate for $\meg$ and the alteration of the sparticle spectrum. In either case we can 
see that the hidden sector has a dramatic effect on the rate for $\meg$ and that a knowledge of the 
hidden sector is not only required to make accurate predictions of the low-energy mass spectrum, but 
is also required to predict the level of expected charged-lepton flavour violation.

\begin{figure}[ht!]
   \newlength{\picwidthf}
   \setlength{\picwidthf}{7in}
   \begin{center}
       \resizebox{\picwidthf}{!}{\includegraphics{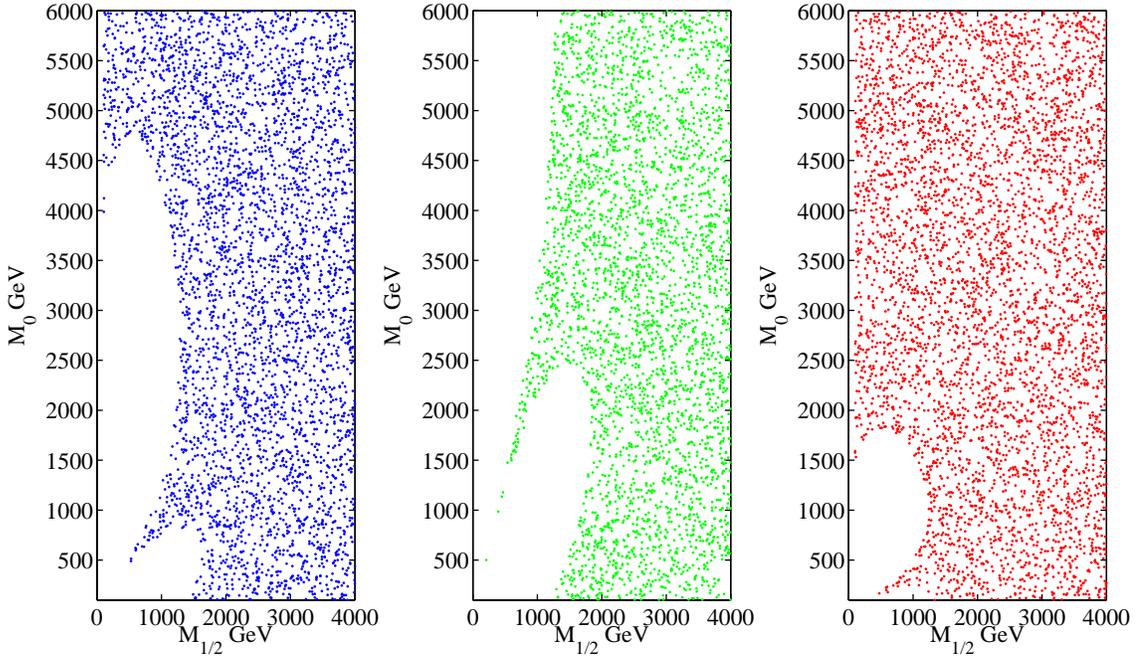}}
   \end{center}
   \caption{\it Degenerate gauge-singlet neutrinos with an IR-free hidden-sector theory,
   displaying the allowed parameter space with $BR(\meg) < 1.2\times 10^{-11}$, with the solid 
   white regions denoting excluded regions. Left panel: allowed parameter space without 
   hidden-sector dynamics, middle panel: allowed parameter region with $b_\gamma = 2/3$, 
   $a_\gamma = \log(5\times 10^{18}\hspace{1mm} \mathrm{GeV}/\mu)$, and right panel: 
   allowed parameter region with $b_\gamma = -2/3$, $a_\gamma = \log(5\times 10^{18}
   \hspace{1mm} \mathrm{GeV}/\mu)$. In all cases $M_R = 10^{14}$~GeV, 
   $M_{\mathrm{hidden}}=10^{12}$~GeV, $\mu >0$, $\tan\beta = 20$, and $a_0=0$.}
   \label{D_IR_free}
\end{figure}

\section{Hidden-Sector Dynamics and Seesaw Reconstruction}

Since the hidden sector affects not only the supersymmetric spectrum but also the radiatively-induced
seesaw charged LFV, the hidden sector impacts the ability to reconstruct the seesaw from low-energy
observations. As we see from eq.(\ref{slepton_L}), charged LFV fixes the elements of 
$\Yn\Yn^\dagger$, while low-energy neutrino observations determine the PMNS matrix and 
$\kappa$ in the Casas-Ibarra parametrization~\cite{Casas:2001sr}, eq.(\ref{CI_par}). It has been 
shown~\cite{Davidson:2001zk} that, in principle, in the seesaw extension of the MSSM, a unique map 
$\Yn\Yn^\dagger, \kappa \rightarrow \Yn, \mathcal{M}$ exists, and thus the seesaw parameters can 
be reconstructed from low-energy data. What is important in this reconstruction is that in the 
seesaw extension of the MSSM all the non-seesaw interactions are in principle known from low-energy
data, so the RGE mixing responsible for the $\Yn\Yn^\dagger$ generating charged LFV
has only the seesaw parameters as unknowns.

In our preceding paper~\cite{CEM1}, we made the case that, for hidden sectors that could be 
effectively parametrized by a two-parameter characterization of the anomalous dimension of the 
dominant scalar mass-squared mediation operator (after removing external line wave-function
renormalization effects by rescaling the input parameters at $M$) that the parameters could in 
principle be fit, and in most cases distinguished from other parametrizations by measurements 
of the low-energy soft supersymmetry-breaking terms. This reconstruction can be done using 
combinations of the soft parameters which are unaffected by the seesaw.

In these circumstances one would still be able to do seesaw reconstruction, as all the non-seesaw contributions to the soft-parameter RGEs would be known. However, the reconstruction does 
depend on the assumption that the slepton soft mass matrices at the mediation scale $M$ are 
proportional to the identity, as in the CMSSM case.

Hence, if the LHC and a linear collider reveal evidence for a distorted superpartner spectrum,
consistent with strong hidden-sector dynamics, it will be essential to reconstruct this dynamics as 
proposed in~\cite{CEM1} first, before one can use other low-energy observables, such as 
charged-lepton flavour violation, to reconstruct the parameters of the seesaw mechanism.

Conversely, if the hidden sector dynamics is sufficiently complicated to defy convenient 
parametrization and experimental determination, then the unknown hidden-sector contributions 
to the RGE flow of the soft supersymmetry-breaking parameters will present an irreducible 
obstacle to using low-energy data to determine seesaw parameters.

\section{Comments and Conclusions}

Recent theoretical observations have demonstrated that low-energy predictions in supersymmetric 
models, such as the supersymmetric spectrum itself, can be significantly influenced by a dynamical 
hidden sector used to break supersymmetry~\cite{Dine:2004dv,Cohen:2006qc,Murayama:2007ge,Schmaltz:2006qs}. In this paper, we have examined the effect of a dynamical hidden sector 
on seesaw induced charged-lepton flavour violation. Since both the seesaw sector and the hidden 
sector sit at intermediate scales, an interval exists where both effects are active during the 
renormalization group running from the unification scale to the weak scale. The combined effect 
may alter the usually expected charged-lepton flavour violation in the seesaw extension of the MSSM.

In our analysis, we parameterized the effect of the hidden sector through a simple 
anomalous-dimension Ansatz. In any realistic model, the actual behaviour of the hidden-sector 
coupling would be determined by the theory, but we see that from our simple parametrization that 
we can capture the effect of moderately-coupled IR-free and UV-free theories. At a comparison 
point, we see that the self-coupling in the cubic superpotential theory requires $\lambda \gtrsim 5$ 
in order for the hidden sector to have a significant effect on the rate for $\meg$. Our 
model-independent parametrizations ensured perturbativity in the range of integration and, in particular, 
we ensured the the hidden sector remains perturbative up to the reduced Planck scale. In order to 
generate a large effect on $\meg$, we need a moderately strongly-coupled hidden sector. In the 
UV-free theory, we placed the pole of the anomalous dimension just below the hidden-sector scale
itself. This ensured that the anomalous dimension becomes large over the range where the seesaw 
is active. If the hidden sector does not become strongly coupled until well outside the interval between 
the hidden-sector and Majorana scales, the effect on flavour violation becomes minimal. We can see 
this effect with the IR-free theory where we place the pole just above the reduced Planck mass. 
Since we started running the theory at $M_X = 2\times 10^{16}$ GeV, the anomalous dimension 
started from a smaller value relative to the end-point of the UV-free anomalous dimension in our 
examples.

Finally, we noted that, in the presence of strongly-coupled hidden sectors whose dynamics
distorts significantly the pattern of soft supersymmetry-breaking scalar masses-squared, the ability to 
use soft parameters in a reconstruction of the seesaw depends on establishing previously
the ability to use the TeV-scale observables to reconstruct the hidden-sector dynamics in ways 
that we have previously analyzed in~\cite{CEM1}. Otherwise, the observable effects of the
hidden sector may impede our view of the seesaw sector.

\section*{Acknowledgements}

We are deeply grateful to Graham Ross for sharing with us the parametrization of the hidden-sector
renormalization effects presented in Section 4, and also for his encouragement during the early 
part of this work. BC and DM would like to acknowledge the support of the Natural Sciences 
and Engineering Research Council of Canada.

\section*{Appendix: One-Loop MSSM Calculation for $l_j \rightarrow l_i + \gamma$}

We calculate the rate for $\meg$ at one loop in the MSSM after running the full system of
MSSM RGEs. We follow the notation in~\cite{Jankowski:2004ve}; similar formulae can be found 
in~\cite{Hisano:1995cp}. The interactions leading to the LFV process $l_j \rightarrow l_i + \gamma$ 
involve the effective Lagrangians describing the neutralino-lepton-slepton and the 
chargino-lepton-sneutrino systems. Written in the mass eigenbasis where $\ON$ diagonalizes 
the neutralino mass matrix, $\OL$ and $\OR$ diagonalize the chargino mass matrix, $\Uf$ 
diagonalizes the charged-slepton mass matrix, and $\Un$ diagonalizes the weak-scale sneutrino 
mass matrix, we have
%
%
%
\begin{equation}
\label{eq-nls}
\la =
\sum_{i=1}^3 \sum_{a=1}^4 \sum_{b=1}^6
N^\mathrm{L}_{iab} \tilde{f}_b E_i \tilde{\chi}^0_a
+ N^{\mathrm{R}*}_{iab} \tilde{f}^*_b e_i \tilde{\chi}^0_a
\cc
\end{equation}
and
\begin{equation}
\label{eq-cls}
\la =
\sum_{i=1}^3 \sum_{a=1}^2 \sum_{b=1}^3
C^\mathrm{L}_{iab} \tilde{\nu}_b E_i \tilde{\chi}^-_a
+ C^{\mathrm{R}*}_{iab} \tilde{\nu}^*_b e_i \tilde{\chi}^+_a ,
\cc
\end{equation}
where
\begin{eqnarray}
\label{eq-NL}
N^\mathrm{L}_{iab} &=&
- \frac{g_2}{\sqrt{2}}
\left(
2\tan\theta_\mathrm{W}
\left(\Uf\right)^*_{b\left(i+3\right)}
\left(\ON\right)_{a1}
+ \frac{m_{\mathrm{l}_i}}{m_\mathrm{W}\cos\beta}
\left(\Uf\right)^*_{bi}
\left(\ON\right)_{a3}
\right),\\
N^\mathrm{R}_{iab} &=&
\frac{g_2}{\sqrt{2}}
\left(
\tan\theta_\mathrm{W}
\left(\Uf\right)^*_{bi}
\left(\ON\right)_{a1}
+ \left(\Uf\right)^*_{bi}
\left(\ON\right)_{a2}
- \frac{m_{\mathrm{l}_i}}{m_\mathrm{W}\cos\beta}
\left(\Uf\right)^*_{b\left(i+3\right)}
\left(\ON\right)_{a3}
\right),\quad \quad
\end{eqnarray} ,
and
\begin{eqnarray}
\label{eq-CL}
C^\mathrm{L}_{iab} &=&
\frac {g_2 m_{\mathrm{l}_i}} {\sqrt{2} m_\mathrm{W} \cos \beta}
\left(\OL\right)_{a2}
\left(\Un\right)^*_{bi},\\
\label{eq-CR}
C^\mathrm{R}_{iab} &=&
-g_2
\left(\OR\right)_{a1}
\left(\Un\right)^*_{bi}.
\end{eqnarray}

The on-shell amplitude for $l_j \rightarrow l_i + \gamma$ has the general form
\begin{equation}
\label{eq-ampl}
\mathcal{M} =
e \epsilon^*_\mu \bar{l}_i\left(p-q\right)
\left(i m_{\mathrm{l}_j} \sigma^{\mu\nu} q_\nu
\left( A_\mathrm{L} \mathrm{L} + A_\mathrm{R} \mathrm{R} \right)
\right)
l_j \left(p \right) ,
\end{equation}
where we have used Dirac spinors
$l_i\left(p-q\right)$ and $l_j\left(p\right)$
for the charged leptons $i$ and $j$ with momenta $p-q$ and $p$,
respectively;
$\mathrm{L}=\left(1-\gamma^5\right)/2$ and
$\mathrm{R}=\left(1+\gamma^5\right)/2$.
Each of the dipole coefficients $A_\mathrm{L}$ and $A_\mathrm{R}$
receives contributions from the neutralino-lepton-slepton
and chargino-lepton-sneutrino interactions, namely,
\begin{equation}
\label{eq-AL}
A_\mathrm{L} = A_\mathrm{L}^\mathrm{(n)} + A_\mathrm{L}^\mathrm{(c)},
\end{equation}
and
\begin{equation}
\label{eq-AR}
A_\mathrm{R} = A_\mathrm{R}^\mathrm{(n)} + A_\mathrm{R}^\mathrm{(c)},
\end{equation}
where
$A_\mathrm{L}^\mathrm{(n)}$,
$A_\mathrm{R}^\mathrm{(n)}$,
$A_\mathrm{L}^\mathrm{(c)}$,
$A_\mathrm{R}^\mathrm{(c)}$
can be evaluated from the Feynman diagrams in Fig.~\ref{loop_dia};
\begin{eqnarray}
\label{eq-ALn}
A_\mathrm{L}^\mathrm{(n)} & = &
\phantom{-}\frac {1} {32 \pi^2}
\sum_{a=1}^{4} \sum_{b=1}^{6}
\frac {1} {m^2_{\mathrm{\tilde{f}}_b}}
\left(
N^\mathrm{L}_{iab} N^{\mathrm{L}*}_{jab}
J_1 \left( \frac {M^2_{\tilde{\chi}^0_a}} {m^2_{\tilde{l}_b}} \right)\right. \nonumber \\
&&\left.+N^\mathrm{L}_{iab} N^{\mathrm{R}*}_{jab}
\frac{ \left|M_{\tilde{\chi}^0_a}\right| } {m_{l_j}}
J_2 \left( \frac {M^2_{\tilde{\chi}^0_a}} {m^2_{\tilde{l}_b}} \right)
\right), \\
%
%
\label{eq-ALc}
A_\mathrm{L}^\mathrm{(c)} & = &
-\frac {1} {32\pi^2}
\sum_{a=1}^{2} \sum_{b=1}^{3}
\frac {1} {m^2_{\tilde{\nu}_b}}
\left(
C^\mathrm{L}_{iab} C^{\mathrm{L}*}_{jab}
J_3 \left( \frac {M^2_{\tilde{\chi}^-_a}} {m^2_{\tilde{\nu}_b}} \right)\right. \nonumber \\
&&+\left. C^\mathrm{L}_{iab} C^{\mathrm{R}*}_{jab}
\frac{ M_{\tilde{\chi}^-_a} } {m_{l_j}}
J_4 \left( \frac {M^2_{\tilde{\chi}^-_a}} {m^2_{\tilde{\nu}_b}} \right)
\right), \\
A_\mathrm{R}^\mathrm{(n)} &=&
\left.A_\mathrm{L}^\mathrm{(n)}\right|_{L \leftrightarrow R}, \\
A_\mathrm{R}^\mathrm{(c)} &=&
\left.A_\mathrm{L}^\mathrm{(c)} \right|_{L \leftrightarrow R}.
\end{eqnarray}
The functions
$J_1\left(x\right)$,
$J_2\left(x\right)$,
$J_3\left(x\right)$,
$J_4\left(x\right)$
are defined as
\begin{eqnarray}
J_1\left(x\right) &=&
\frac
{1 - 6x + 3x^2 + 2x^3 - 6x^2 \ln x}
{6\left(1-x\right)^4},\\
J_2\left(x\right) &=&
\frac
{1 - x^2 + 2x\ln x}
{\left(1-x\right)^3},\\
J_3\left(x\right) &=&
\frac
{2 + 3x - 6x^2 + x^3 + 6x \ln x}
{6\left(1-x\right)^4},\\
J_4\left(x\right) &=&
\frac
{-3 + 4x - x^2 + 2\ln x}
{\left(1-x\right)^3}.
\end{eqnarray}
Finally, the decay rate for $l^-_j \rightarrow l^-_i + \gamma$ is given by
\begin{equation}
\label{eq-ratio}
\Gamma\left(l^-_j \rightarrow l^-_i + \gamma\right) =
\frac {e^2} {16\pi} m_{l_j}^5
\left(
\left|A_\mathrm{L}\right|^2
+\left|A_\mathrm{R}\right|^2
\right),
\end{equation}
and $i=1$, $j=2$ for $\mu \rightarrow e + \gamma$.

%
%
%


 





\end{document}